\documentclass[english,aps,prl,twocolumn,showpacs,groupaddress,floatfix,graphics,graphicx]{revtex4-1}
\usepackage[latin9]{inputenc}
\usepackage{amsmath}
\usepackage{amssymb}
\usepackage{graphicx}
\usepackage{verbatim}
\usepackage{braket}
\usepackage{soul}
\usepackage[usenames]{color}
\usepackage{tikz}
\usetikzlibrary{matrix}

\renewcommand\vec{\boldsymbol}

\usepackage{babel}

\begin{document}

\title{Scattering of magnons at graphene quantum-Hall-magnet junctions}

\author{Nemin Wei, Chunli Huang, and Allan H. MacDonald}

\affiliation{Department of Physics, The University of Texas at Austin, Austin,
Texas 78712, USA}

\begin{abstract}
Motivated by recent non-local transport studies of quantum-Hall-magnet (QHM) states formed in monolayer graphene's $N=0$ Landau level, 
we study the scattering of QHM magnons by gate-controlled junctions between states with different integer filling factors $\nu$.
For the $\nu=1|-1|1$ geometry we find magnons are weakly 
scattered by electric potential variation in the junction region, and that 
the scattering is chiral when the junction lacks a mirror symmetry.  
For the $\nu=1|0|1$ geometry, 
we find that kinematic constraints completely block magnon transmission if the incident angle exceeds a critical value.
Our results explain the suppressed non-local-voltage signals observed in the $\nu=1|0|1$ case.  
We use our theory to propose that valley-waves generated at $\nu=-1|1$ junctions and magnons can
be used in combination to probe the spin/valley flavor structure of QHM states at integer and fractional filling factors.
\end{abstract}

\maketitle

\textit{Introduction}-- 
The recent discovery of magnetic order in two-dimensional materials \cite{samarth2017condensed, gibertini2019magnetic,burch2018magnetism,huang2018electrical,jiang2018controlling} 
has suggested new strategies to build ultra-compact spintronic devices that utilize magnons as weakly dissipative information 
carriers \cite{chumak2015magnon,chumak2019fundamentals,rezende2020magnon}. %
Ordered states, referred to generically as quantum Hall magnets (QHMs), occur in graphene in a strong magnetic field
and break spin and valley symmetries
\cite{zhang2006landau,checkelsky2009zeroenergy,du2009fractional,young2012spin,young2014tunable,goerbig2011graphene,goerbig2006graphene,
alicea2006iqhe,yang2006grapheneQHF,doretto2007qhf,herbut2007iqhe,jung2009QHF,nomura2009KT,
nandkishore2012bilayer,kharitonov2012bilayer,kharitonov2012phase,sodemann2014fractional}.
Because of their electronic simplicity and gate tunability,
and also because the technology needed to prepare extremely clean and well characterized monolayer graphene samples is 
well established \cite{dean2020fractional, dean2010boron,dean2011multicomponent,zibrov2017tunable}, 
graphene QHMs are an excellent system in which to demonstrate
two-dimensional spintronic and magnonic device concepts.

When a strong magnetic field is applied perpendicular to a 2D graphene sheet,
the $\pi$-orbitals of the carbon atoms form Landau levels with approximate four-fold isospin degeneracy.
The isospin degeneracy combines a two-fold valley pseudospin with the electron spin degree of freedom.
In a partially filled Landau level, Coulomb interactions often break the
Hamiltonian's SU$(4)$ isospin symmetry and give rise to a rich family of 
correlated insulating states. 
At an integer filling factor, the ground-state 
is a single Slater determinant and can be therefore described by Hartree-Fock mean-field theory \cite{girvin1999quantum,nomura2006QHF,kharitonov2012phase}.
At filling factor $\nu=\pm 1$, {\it i.e.} at three-quarter and one quarter-filling of the $N=0$ Landau level quartet, 
the ground-state is analogous to the QHM states found in two-dimensional electron gases in 
semiconductor quantum wells and consists of fully spin and valley polarized electrons ($\nu=-1$) or holes ($\nu=1$)\cite{abanin2013fractional}.
In contrast, the ground state at filling factor $\nu=0$ (half-filling of the $N=0$ Landau level) is more complicated.
As pointed out by Kharitoniv \cite{kharitonov2012phase}, the $\nu=0$ phase-diagram contains a ferromagnet(F), 
a canted antiferromagnet (CAF), a Kekul\'{e} distortion state, and a charge density wave. 
The competition between these states is influenced by weak lattice-scale Coulomb interactions that break SU$(4)$ symmetry, 
sample-dependent substrate-induced sublattice polarization potentials\cite{hunt2013massive,amet2013insulating,zibrov2018even},
dielectric screening \cite{veyrat2020helical} and in-plane magnetic fields. The systematic \cite{young2014tunable} dependence on in-plane magnetic field of an edge-state metal-insulator transition
strongly suggests that the $\nu=0$ ground state is a canted antiferromagnet (CAF) in which opposite valleys have different spin polarizations.  
The ordered states at $\nu=0,\pm 1$ support low-energy collective excitations   
 \cite{alicea2006iqhe,yang2006grapheneQHF,doretto2007qhf} that are analogous to magnon modes in a conversational magnetic systems, and 
 which we will refer to as QH magnons.

Recent experiments \cite{stepanov2018long,wei2018electrical,zhou2019skyrmion} have studied
the transmission of QH magnons through junctions between distinct QHM states.
In Ref.~\cite{wei2018electrical,zhou2019skyrmion}, $\nu=1$ QH magnons are generated electrically by driving magnon-mediated transitions between conducting edge 
states with different spin-orientations.  The change in conduction spin is transferred to 
a magnon that can be propagated through the two-dimensional bulk. (See Ref.~\cite{huang2020qhsj} for a theoretical model of the 
magnon generation process.) Magnons are then guided toward $1|\nu_{m}|1$ QHM junctions, where 
$\nu_{m}$ is a (gate-tunable) filling fraction of interest sandwiched between $\nu=1$ regions. 
Any magnons transmitted through the junction generate non-local electrical signals on the opposite side of the 
device via the reciprocal of the magnon generation process.
Measured non-local voltages suggest that the $1|-1|1$ junction is nearly transparent for magnons, since 
the non-local voltage signal is not greatly reduced by its presence.  
In contrast, the non-local voltage signal is greatly suppressed by $\nu= 1|0|1$ junctions.
This finding requires an explanation since the $\nu_m=0$ canted antiferromagnet also 
supports magnons\cite{murthy2016modes,denova2017collective,pientka2017thermaltransport}.

In this Letter, we use microscopic theory to calculate magnon transmission through $1|\nu_m|1$ QHM junctions. 
For $\nu_{m}=-1$ we find that although the magnon modes are identical in all regions, the  
electrostatic inhomogeneity of the junction partially reflects magnons.
The $\nu_m=0$ CAF state has two magnon branches that, except at very small momenta, have higher energies than $\nu=1$ magnons.
We find that this energy mismatch leads to perfect reflection above a critical angle of incidence $\Theta_c$, explaining the difference
in non-local electrical signals.  

\begin{figure}[t]
\includegraphics[width=1\columnwidth]{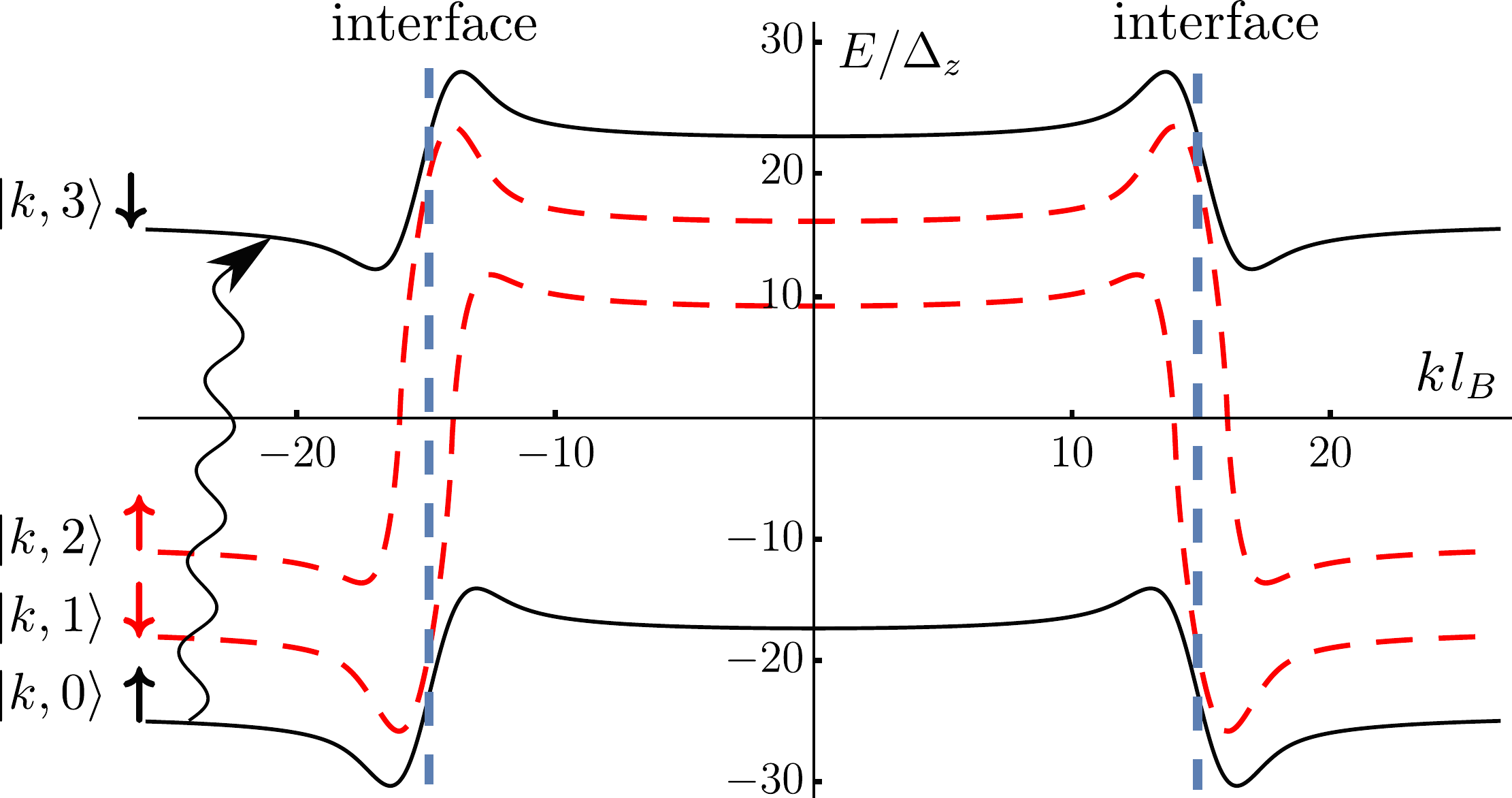}
\caption{ Self-consistent Hartree-Fock bandstructure of a $1|-1|1$ junction in which the 
sense of valley polarization is opposite in the $\nu=1$ and $\nu=-1$ regions.  The uniform 
$\nu= 1$ and $\nu=-1$ states have majority ($\uparrow$) spin occupation selected by the 
weak Zeeman coupling and, for unaligned hBN encapsulation, spontaneously chosen valley polarization. 
The black solid lines show valley $K$ quasiparticle energies {\it v.s.~}guiding center, and the 
red dashed lines show the valley $K'$ orbitals 
that cross the Fermi level ($E_F=0$) at $\nu=1|-1$ junctions. The curly line represents the bands involve in particle-hole transition of a $\nu=\pm1$ magnon.
  }\label{fig:1-11bandstructure}
\end{figure}%

\textit{Time Dependent Hartree-Fock Theory}-- We formulate the problem of collective-mode transmission by studying the 
dynamics of the $N=0$ Landau level single-particle density-matrix
\begin{equation}
i\partial_{t}\hat{P}(t)=[\hat{H},\hat{P}(t)], \label{eq_vonneuman}
\end{equation}
where $\hat{H}$ is the mean-field Hamiltonian determined self-consistently at each instant in time:
\begin{subequations}
\begin{align}
\hat{H}_{k+q_y,k}&=\hat{H}^0_k \delta_{q_{y,0}}+\hat{\Sigma}^{H}_{k+q_{y},k}+\hat{\Sigma}^{F}_{k+q_{y},k},\\
\hat{H}^{0}_k&=\frac{\Delta_z}{2}s^{z}+ \frac{\Delta_{v}}{2}\tau^{z}+E_b(k),\label{eq_H0}\\
\hat{\Sigma}^{H}_{k+q_{y},k}&=\sum_{\alpha=0}^{3}\sum_{k'}V_{\alpha}(k-k',q_{y})\text{tr}(\tau^{\alpha}\hat{P}_{k'+q_{y},k'})\tau^{\alpha},\label{eq_Hartree}\\
\hat{\Sigma}^{F}_{k+q_{y},k}&=-\sum_{\alpha=0}^{3}\sum_{k'}V_{\alpha}(q_{y},k-k')\tau^{\alpha}\hat{P}_{k'+q_{y},k'}\tau^{\alpha}.\label{eq_Fock}
\end{align}
\end{subequations}
The single-particle Hamiltonian $\hat{H}^0_k$, specified in Eq.~\eqref{eq_H0}, 
includes Zeeman energy ($\Delta_z=\mathfrak{g}\mu_B |B|$), valley-polarization energy  ($\Delta_v$)
and background electrostatic ($E_b$) energy contributions.  $\Delta_v$ is induced by adjacent hBN layers if 
aligned and $E_b$ controls the spatial variation of filling fraction.
Here $\vec{s}$ ($\vec{\tau}$) are Pauli matrices in spin (valley) space and the wavevectors $k$ are Landau gauge momenta in the direction 
along the junction line.  The electrostatic background potential $E_b$ is $k$ dependent because Landau gauge eigenstates are 
localized along guiding center lines with $x$-coordinate $X=kl_{B}^2$, where $l_B$ is the magnetic length.
In Eqs.~\eqref{eq_Hartree}--\eqref{eq_Fock}, the $\alpha=0$ and $\alpha=1,2,3$ self-energy terms
account respectively for the SU$(4)$ invariant long-range Coulomb interaction and the short-range valley-dependent interactions.\footnote{See Supplementary Material for the details of the model, the properties of the collective modes and the numeric methods to calculate transmission probability of collective modes.} 
The time-independent self-consistent solutions of Eq.~\eqref{eq_vonneuman} preserves translational symmetry along the junction line and is therefore diagonal in $k$\cite{wei2020band}:
\begin{equation}
 \hat{P}_{k+q_{y},k}^{0}=\delta_{q_{y},0}\sum_{m=0}^{3} f_{m,k}\ket{k,m}\bra{k,m},
\end{equation}
where $\ket{k,m}$ is the $m$-th mean-field band ordered energetically from $0$ to $3$ and 
$f_{m,k}$ is its occupation number.  We plot the quasiparticle bandstructure of a $\nu=1|-1|1$ junction in Fig.~\ref{fig:1-11bandstructure} for future reference. 
To describe small amplitude dynamics, we expand $\hat{P}(t)=\hat{P}^{0} + \delta \hat{P}(t)$ and use the compact notation 
\begin{equation} \label{eq:psi_kmn}
\psi_{kmn}(q_{y}) \equiv  \langle k+q_y,m| \delta\hat{P}|k,n\rangle,
\end{equation}
to denote particle-hole transition amplitudes with momentum $q_y$.
When linearized in $\delta\hat{P}$, Eq.~\eqref{eq_vonneuman} implies that 
\begin{equation}
\omega\, \psi_{kmn}(q_{y},\omega)=\sum_{k'm'n'}\mathbb{K}_{kmn}^{k'm'n'}(q_{y})\,\psi_{k'm'n'}(q_{y},\omega),\label{eq_RPA}
\end{equation}
where $\omega$ is the collective mode frequency and $\mathbb{K}_{kmn}^{k'm'n'}$ is known as the RPA (random-phase approximation 
\cite{negele1982tdhf,ring2004nuclear,murthy2016modes,denova2017collective,pientka2017thermaltransport} ) kernel
that acts as a superoperator on the collective mode $\psi$\cite{Note1}.

%
%

\begingroup
\squeezetable
\begin{table}
    \centering
    \caption{
    Properties of the magnon mode $\omega_s$ of the $\nu=1$ F state and the two magnon modes 
    $\omega_{1,2}$ of the  $\nu=0$ CAF state.  The CAF  modes are linear-combinations of spin-flips in the $K$ and $K'$ valleys
    and capture the quantum dynamics of the N\'{e}el $\vec{n}$ and spin-polarization $\vec{m}$ vectors.
    $\phi_{mn}(\vec{q})\equiv\sum_{k} e^{iq_{x}kl_{B}^2}\psi_{kmn}(q_{y})$. \cite{Note1}}
    \begin{ruledtabular}
    \begin{tabular}{l l l l}
    &$(\phi_{30},\phi_{21},\phi_{12},\phi_{03})$&Gap&\textrm{Description}\\
    \hline
    $\omega_{s}$&$(1,0,0,0)$&$\Delta_{z}$&spin precession within a  valley\\
    $\omega_{1}$&$(u_{1\vec{q}},u_{1\vec{q}},v_{1\vec{q}},v_{1\vec{q}})$&0&in-plane($\perp\vec{B}$) oscillation of $\vec{n}$\\
    $\omega_{2}$&$(u_{2\vec{q}},-u_{2\vec{q}},v_{2\vec{q}},-v_{2\vec{q}})$&$\Delta_{z}$&precession of $\vec{m}$ about $\vec{B}$ field\\
    \end{tabular}\\
    \end{ruledtabular}
    \label{tab:modes}
\end{table}
\endgroup

\begin{figure*}
\includegraphics[width=2\columnwidth]{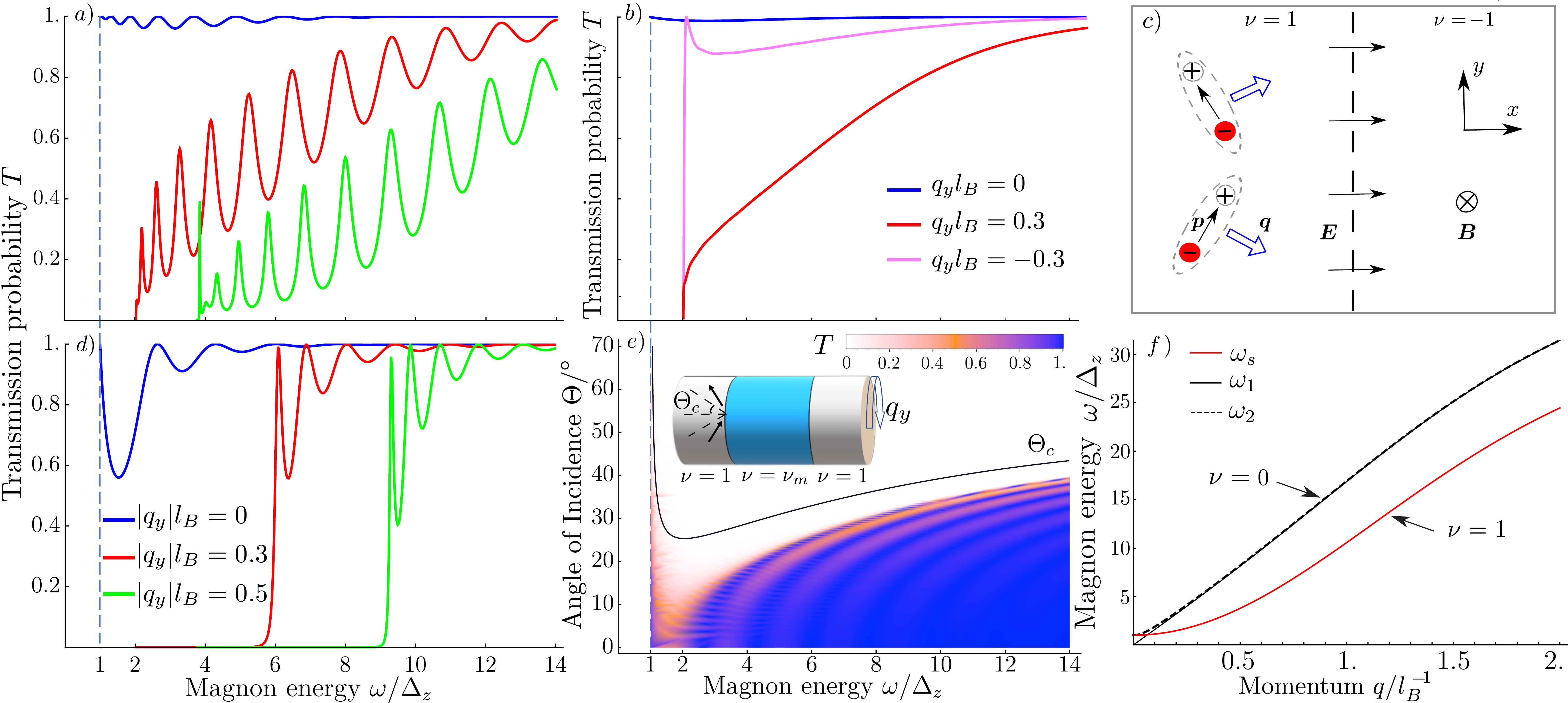}
\caption{
Magnon transmission probabilities $T(q_y,\omega)$ v.s.~$\omega$ for $\nu=1|-1|1$ (a), $\nu=1|-1$(b) and $\nu=1|0|1$ (d) QHM junctions\cite{Note1}. 
%
c)  Schematic particle-hole pairs in $\nu=1|-1$ junctions. 
The interfacial electric field $\vec{E}$ points from $\nu=1$ to $\nu=-1$. 
Negative (positive) signs represents electrons (holes). 
The dipole moment $\vec{p}$ of electron-hole pairs is perpendicular to both the magnetic field $\vec{B}$ and the center-of-mass 
momentum $\vec{q}$.
e) Color plot of the magnon transmission probability through a $\nu=1|0|1$ junction {\it vs.} energy and angle of incidence. 
f) Magnon dispersions in uniform $\nu=\pm 1$ F states ($\omega_s$) and in $\nu=0$ ($\omega_{1,2}$) CAF states.
These results are generated with experimental determined Coulomb interaction strength at $B=8T$ in a geometry with width $L_y=80\pi l_B$ and the length of $\nu_m$ region is $30l_{B}$.
}\label{fig:singlemodeT}
\end{figure*}
%

\textit{Magnon Scattering}-- The magnon scattering problem is complicated by the strong non-locality of the RPA kernel $\mathbb{K}_{kmn}^{k'm'n'}(q_{y})$.
In the absence of a junction $\mathbb{K}_{kmn}^{k'm'n'}(q_{y})$ is invariant under simultaneous translation of guiding centers
$kl_B^2$ and $k'l_B^2$, allowing Eq.~\ref{eq_RPA} to be solved by Fourier transformation to obtain bulk modes labelled by two-dimensional 
wavevectors $\mathbf{q}=(q_x,q_y)$ with energies $\omega_i(\mathbf{q})$.
Some key properties of the bulk collective modes are briefly summarized in Table.~\ref{tab:modes}.
Since $q_y$ remains a good quantum number in the presence of a  $1|\nu_m|1$ junction, we are left with a $q_y$-dependent one-dimensional scattering problem
with the $\nu=1$ bulk modes as asymptoptic states.  We therefore apply the scattering boundary conditions:
\begin{align}   
&\psi_{k30}(q_{y},\omega)=
\begin{cases}
e^{iq_{x}kl_{B}^2}+r(q_y,\omega)\,e^{-iq_{x}kl_{B}^2},&k\rightarrow -\infty\\
t(q_y,\omega) \, e^{iq_{x}kl_{B}^2},&k\rightarrow\infty
\end{cases}\notag\\
&\psi_{kmn}(q_{y},\omega)=0,\quad k\rightarrow \pm \infty\textrm{ and } m,n\neq (3,0)\label{def_bcs}.
\end{align}
The asymptotic states are pure ($\psi_{k30}$) $\nu=1$ magnons that are gapped by the Zeeman energy \cite{wei2018electrical,zhou2019skyrmion}.
In Eq.~\eqref{def_bcs} $q_x$ is determined by solving $\omega_s(\mathbf{q})=\omega$.
We solve for the scattering states and the $q_y$-dependent reflection $r(q_y,\omega)$ and transmission $t(q_y,\omega)$ coefficients by 
discretizing $k$, applying Eq.~\ref{eq_RPA} at $j=1,...N$ points in a scattering region centered on the junction, and substituting the asymptotic expressions for 
$\psi_{k'm'n'}(q_{y},\omega)$ at $j=1$, $j=N$, and outside the junction.  Only the ${m,n}=(3,0)$ RPA equation is applied at $j=1$ and $j=N$, which are assummed to be 
in the asymptotic region.  This procedure yields a set of inhomogeneous linear equations \cite{Note1} that we 
have converged with respect to guiding center mesh density to obtain the results discussed below.

\textit{Magnon Transmission Results}-- Our results for the magnon transmission probabilities $T(q_y,\omega)=|t(q_y,\omega)|^2$ of 
$1|\nu_m|1$ QHM junctions with $\nu_m=-1$ and $\nu_m=0$ are shown in Figs.~\ref{fig:singlemodeT}a) and d) respectively. 
Both junctions have a threshold energy $\omega_{tr}$, below which there is no transmission, $T(q_y,\omega<\omega_{tr})=0$.
For a $1|-1|1$ junction, the bulk $\nu=\pm1$ regions have identical magnon dispersions, so the threshold energy is simply the bulk magnon 
energy at normal incidence: $\omega_{tr} = \omega_s(0,q_y)$.  For $\omega > \omega_{tr}$, we find magnon transmission decreases with increasing $q_{y}$. 
The reduction is due to a peculiar property of collective mode excitations in quantum Hall systems, 
namely that the centre-of-mass momentum $\vec{q}$ of a particle-hole excitation is related to its electric-dipole moment $\vec{p}$ by 
\cite{gor1968contribution,kallin1984excitation,cao2020quantum}, $\vec{p}=|e|l_{B}^2\hat{\vec{z}}\times\vec{q}$, as illustrated in Fig.~\ref{fig:singlemodeT}c).
Magnons with larger $q_{y}$ scatter more strongly off the electric fields $E\hat{x}$ present in the junction region.
When we examine the $1|-1$ and $-1|1$ junctions separately, we find that magnons with opposite signs of $q_y$ have different transmission 
probabilities, as shown in Fig.~\ref{fig:singlemodeT}b).
This behavior is expected since the $1|-1$ junction acts like a repulsive scatterer 
when the dipole moment has an $\hat{x}$ projection opposite to the the junction electric field, 
and like an attractive scatterer when the $\hat{x}$ projection 
has a dipole moment that is aligned with the junction electric field.  The total transmission through 
the $1|-1|1$ junction plotted in Fig.~\ref{fig:singlemodeT}a) and d) has $q_y \to - q_y$ symmetry
because the studied model has mirror symmetry about the $y-z$ plane at the center of the $\nu=\nu_{m}$ region.
We have verified that the junction scattering becomes chiral when this symmetry is absent.

The threshold energy $\omega_{tr}$ in Fig.~\ref{fig:singlemodeT}d ($1|0|1$ junction) appears to be significantly larger
than in Fig.~\ref{fig:singlemodeT}a ($1|-1|1$ junction).  
The suppressed magnon transmission 
is due to a mismatch between CAF and F collective mode dispersions.
As shown in Fig.~\ref{fig:singlemodeT}f), the bulk collective modes of $\nu=0$ CAFs disperse 
more strongly than those of $\nu=1$ Fs, so that $\omega_{1,2}$ has higher energy than $\omega_s$, 
except at very small momenta where $\omega_1$ is gapless while $\omega_2$ and $\omega_s$ are gapped. 
To transmit a $\nu=1$ magnon with energy $\omega=\omega_s$ and parallel momentum $q_y$ through $1|0$ junction, the conservation of 
energy and parallel momentum requires that
%
\begin{equation}
 \omega_s\left(q_{x}^{L},q_{y}\right)=\omega_{1}(q_{x}^{R},q_{y}),\label{eq_constraint}
\end{equation}
where $q^{L/R}_x\geq 0$ are the asymptotic normal momenta on the left (L) and right (R) sides of the $1|0$ junction. 
We identify the threshold energy as $\omega_{1}(0,q_{y})$, the value of $\omega$ for which $q_x^R \to 0$.
Since $\omega_{1}(0,q_{y}) > \omega_{s}(0,q_{y})$ we conclude that the $1|0|1$ junction has a higher threshold energy than $1|-1|1$ junction.
Once the incoming magnon energy
exceeds $\omega_{tr}(q_y)$, as illustrated in Fig.~\ref{fig:singlemodeT}d, $T$ rapidly approaches $1$. This property can be understood by noting the valley polarization of
superpositions of $\omega_1$ and $\omega_2$ modes vary on the long length scale $\lambda_0= 2\pi/(q_{x1}^{R}-q_{x2}^{R})$,
where $q_{x1}^R$ and $q_{x2}^R$ are the nearly identical local $x$ wavevectors of the 
nearly degenerate (Fig.~\ref{fig:singlemodeT}f) $\omega_{1,2}$ modes.  A $\nu=0$ magnon can 
therefore maintain the valley polarization of the $\nu=1$ magnon across the junction, provided that the $\nu_{m}$ region
is shorter than $\lambda_0$.
%
Our results for $1|0|1$ junction magnon transmission are summarized in Fig.~\ref{fig:singlemodeT}e),
in which the transmission probability is plotted as a function of energy and angle of incidence $\Theta=\arctan(q_y/q_x^L)$. 
The black curve shows the critical incident angle $\Theta_c$, obtained by solving Eq.~\eqref{eq_constraint} with $q_x^R=0$.
For higher angles of incidence, momentum and energy conservation imply that the magnons are 
evanescent waves in the $\nu=0$ region. 
\begin{figure*}[t]
\includegraphics[width=2\columnwidth]{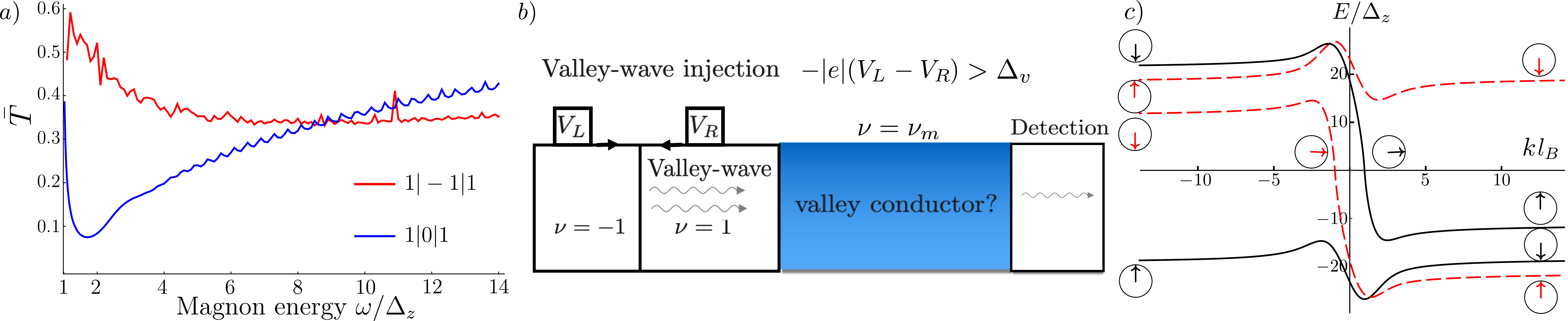}
\caption{ 
 a) Angularly average magnon transmission $\bar{T}(\omega)$ v.s.~$\omega$. 
 The parameters used in this calculation are the same as those in Fig.~\ref{fig:singlemodeT} 
 b) Valley wave scattering devices. We propose to replace the $2|1$ junction used in 
  Ref.~\cite{wei2018electrical,zhou2019skyrmion} with $-1|1$ junctions to generate valley waves.
 c) Bandstructure of a $-1|1$ junction used for valley-wave injection.  
 All states color-coded with black and red are respectively fully polarized in $K$ and $K'$ valleys, 
 while the spin rotates smoothly from $\uparrow$ to $\downarrow$ across the junction. The parameters for this calculation ($B=8T$ and valley polarization energy $\Delta_{v}=3.7$meV 
 correspond to the circumstances of Ref.~\cite{zhou2019skyrmion}).
  }\label{fig:proposal}
\end{figure*}
%
%


The transmission probabilities in Fig.~\ref{fig:singlemodeT} exhibit Fabry-P\'{e}rot oscillations 
generated by the repeated scattering at the two interfaces. 
The interference pattern will be smeared out in observables that average magnons over angles of incidence.
Assumming that all angles of incidence are equally likely, we define an average magnon transmission probability
\begin{equation}
    \bar{T}(\omega) \equiv \frac{1}{\pi}\int_{-\pi/2}^{\pi/2}d\theta\ T(q_y(\theta,\omega),\omega).
\end{equation} 
As shown in Fig.~\ref{fig:proposal}a), the average transmission $\bar{T}$
through a $1|0|1$ junction is noticeably smaller than the transmission through a $1|-1|1$ junction 
at low energies but becomes comparable to a $1|-1|1$ junction at high energy. 
In our calculation of $1|0|1$ junction we assumed perfect screening of induced Hartree potentials in the junction region by nearby 
gates~\cite{wei2020band}. Since the inhomogeneity of the electrostatic potential is a source of magnon-reflection, the transmission through a $1|0|1$ junction 
would be even lower if we accounted for imperfect screening. 

\textit{Discussion:--} 
We now use our findings to interpret the experimental results in Ref.~\cite{wei2018electrical} and to 
propose related studies that might be informative.
Magnons can be generated electrically by bringing edge channels with opposite spins and different 
chemical potentials together at a hot spot, opening a path for magnon-generation mediated edge-channel spin flips.
The energies of magnons generated in this way must be smaller than the electrical bias voltage.
We assume~\cite{huang2020qhsj} that the steady state established by electrically injected magnons~\cite{wei2018electrical},
can be characterized by a magnon distribution with a well defined local chemical potentials.
Non-local voltages generated by the reciprocal of the injection process measure local magnon chemical potentials.
Non-local voltages measured at points that are separated from the injection point by a $1|0|1$ junction, are 
small even when the electrical bias voltages is $\sim 5$ times \cite{wei2018electrical} larger than $\Delta_z$.
This behavior is explained by the larger energies of magnons in $\nu=0$ regions compared to $\nu=1$ regions, as explained above.
The slow increase in average transmission probability $\bar{T}$ with magnon energy we find is also in agreement with experimental trends.
We do find that a peak in $\bar{T}$ (\textit{c.f.}~Fig.~\ref{fig:proposal}a) in a narrow window of energy
( $1<\omega/\Delta_z<1.2$ ) just above $\Delta_z$ where the $\nu=0$ and $\nu=1$ magnon energies are more similar that 
is not detected experimentally, presumably because magnon generation in this energy window is not sufficient 
to produce an observable signal.

The experimental non-local signals of $1|1|1$ and $1|-1|1$ junctions are similar for bias voltages $ \lesssim 5\Delta_z$, 
and much larger than the voltages measured in the $1|0|1$ case.  In our theory this property is due to the fact that $\nu=1$ and 
$\nu=-1$ magnon modes have identical dispersions and therefore no kinematic transmission constraints. 
Our theory does predict finite reflection at $1|-1|1$ junctions that is absent in the translationally invariant $1|1|1$ case, but
this will not be observable if unintended scattering from disorder or the split gate junctions dominates magnon scattering.
Indeed, as we have emphasized, our calculation has identified the electrical dipole moments of QH magnons 
as a mechanism for magnon scattering off variations in electrical potential.  
Other extrinsic mechanisms such as spin-dependent disorder \cite{tikhonov2016emergence,PhysRevB.97.235402,PhysRevB.92.155124} near the sample edges
can also suppress magnon transmission but are unlikely to play a dominant role in high quality devices used in Refs.~\cite{wei2018electrical,zhou2019skyrmion}


In closing we propose an experimental protocol illustrated schematically in Fig.~\ref{fig:proposal}b) to electrically detect valley ordering, e.g.~Kekul\'{e} distortion, by measuring valley-wave transmissions. To inject valley-waves, we replace the $2|1$ interface \cite{wei2018electrical} 
used for magnon-injection with a $-1|1$ interface.
As shown in Fig.~\ref{fig:proposal}c), when the $-1|1$ interface receives finite valley polarization potential from the aligned hBN, the mean-field bandstructure hosts two edge states with opposite valley polarization
and nearly parallel spins whose chemical potentials can be independently controlled via the contacting geometry illustrated in Fig.~\ref{fig:proposal}b).  The bias voltage opens up a path for valley-wave generation scattering between edge channels.
In order to increase valley-wave emission probability, the edge states can be brought into close proximity via a quantum point contact. 
We expect the emitted valley-waves to be transmitted through ground states that support valley-wave excitations.
Measuring non-local voltages provides a new method to determine the isospin structure of quantum Hall ground-states,
which remains an elusive target especially at fractional filling factors \cite{zibrov2018even}.

\textit{Acknowledgement:--}
We acknowledge helpful interactions with Hailong Fu, Andrea Young, Haoxin Zhou and Jun Zhu. This work is supported by DOE BES grant DE- FG02-02ER45958 and by Welch foundation grant TBF1473.
N.W was partially supported by a Graduate School Continuing Fellowship.
\bibliography{ref-DWQH}


\pagebreak
\widetext
\begin{center}
\textbf{\large Supplementary Materials}
\end{center}
\setcounter{equation}{0}
\setcounter{figure}{0}
\setcounter{table}{0}
\setcounter{page}{1}
\makeatletter
\renewcommand{\theequation}{S\arabic{equation}}
\renewcommand{\thefigure}{S\arabic{figure}}
\renewcommand{\bibnumfmt}[1]{[S#1]}
\renewcommand{\citenumfont}[1]{S#1}

\maketitle

\section{ Quantum Hall Magnets: Mean field and collective modes}
We give a systematic introduction to the collective modes in graphene quantum Hall magnet at integer filling fraction of the $N=0$ Landau levels, \textit{i.e.} $\nu=0,1$. Some of the results we discussed here can also be found in literature, see Ref.~\cite{murthy2016modes,denova2017collective,pientka2017thermaltransport}.
We first review the microscopic Hamiltonian projected onto $N=0$ Landau level and its mean-field ground state at various filling fractions. Next, we study collective excitation of the mean-fields using time-dependent Hartree-Fock theory. This is a \textit{conserving approximation} that conserve the symmetries of the microscopic Hamiltonian. Mathematically, we solve the so-called RPA equation whose roots describe the dispersion of collective modes. The results are summarized in Table.~\ref{tab:dispersion} and we neglect $\nu=\pm2$ since they do not support intra-Landau level collective modes.

The microscopic Hamiltonain projected onto the $N=0$ Landau-Level is given by the following:
\begin{equation} \label{eq_H}
\mathcal{\hat{H}}=\sum_{k}c_{k}^{\dagger}\, \hat{H}_{k}^0\, c_{k}+\frac{1}{2}\sum_{\alpha=0}^{3} \,\, \sum_{k,k',q_{y}} V_{\alpha}(k-k',q_{y}):[ c^{\dagger}_{k+q_{y}}\tau^{\alpha}c_{k} ] [c^{\dagger}_{k'}\tau^{\alpha}c_{k'+q_{y}}]:   \; .
\end{equation}
The basis $c_{k}=(c_{kK\uparrow},c_{kK\downarrow},c_{kK'\uparrow},c_{kK'\downarrow})^{T}$ has 4 components in valley ($\vec{\tau}$) and spin ($\vec{s}$) space. The single-particle term is independent of $k$ unless translation symmetry is broken (\textit{e.g.~}close to the edge or domain wall). We set the background potential $E_b(k)=0$ in this section to discuss collective modes in the bulk, then $\hat{H}_{k}^{0}$ consists of the spin-splitting from the Zeeman effect ($\Delta_z$) and a possible valley splitting from the sublattice polarization potential:
\begin{equation}
    \hat{H}_{k}^0= =-\frac{\Delta_z}{2}s^z-\frac{\Delta_v}{2}\tau^z.
\end{equation}
 The second term in Eq.~\eqref{eq_H} describes Coulomb interaction between particles in the $N=0$ Landau level. The Coulomb scattering amplitude is given by the following: 
\begin{equation}
    V_{\alpha}(k-k',q_{y})=\frac{1}{2\pi L_{y}} \int dq_{x}\,U_{\alpha}({\vec{q}})e^{-{\vec{q}^2l_{B}^2}/2}e^{iq_{x}(k-k')l_{B}^2}
\end{equation}
where we use $\alpha=0,1,2,3$ and $\alpha=0,x,y,z$ interchangeably,
\begin{equation}
    U_{0}(\vec{q})=\frac{2\pi e^2}{\epsilon\sqrt{\vec{q}^2+\kappa^2}} 
    \;\;, \;\;
    U_{i}(\vec{q})=2\pi l_{B}^2u_{i},\,i=x,y,z
\end{equation}
 Here $U_{0}$  is the long-range Coulomb potential and it has a infrared cut-off $\kappa\sim1/L_{y}$  and we take the dielectric constant $\epsilon\approx 6.6$ from  a Boron-Nitride substrate. Besides the long-range Coulomb interaction, the short-range valley anisotropic interaction ($U_{i}(\vec{q})$) in graphene is also important in selecting the correct ground-states, as pointed out by Kharitonov \cite{kharitonov2012phase,kharitonov2012bilayer}.  This is because the short-range interaction reduce the $SU(4)$ symmetry of the $U_0$ Hamiltonian. 
 Although momentum non-conserving (\textit{i.e.}~Umklapp) scattering process is allowed by the magnetic field, it is exponentially suppressed by a factor $e^{-(l_{B}/a)^2}$ where $a$ is the lattice constant. So to a very good approximation,
 $u_{x}=u_{y}\equiv u_{\perp}$ and the resulting interacting Hamiltonian has an $U(1)_{v}$ symmetry. Furthermore, the experimental observation of metal-insulator phase transition of the edge states \cite{young2014tunable,kharitonov2012phase} have narrow down the relevant parameter space to 
 \begin{equation} \label{eq:heir}
     0<-u_{\perp}<u_z
 \end{equation}  
In all of our numeric calculations, we use valley anisotropic energies inferred from experiments\cite{young2014tunable,zibrov2018even}: $\ u_{\perp}=-4\Delta_{z}$ and $u_{z}=7\Delta_{z}$ in the perpendicular magnetic field.

\subsection{Mean-Field Ground State}\label{sec:meanfield}

We seek ground-state of Eq.~\eqref{eq_H} with the following mean-field order parameter
\begin{equation}
   P_{ki,k'j}^{0} = \langle \Psi_{\mathrm{QHM}}| c_{ki}^{\dagger} c_{k'j} |  \Psi_{\mathrm{QHM}} \rangle 
\end{equation}
Here $\Psi_{\mathrm{QHM}} $ is the Slater determinant ground-state to be determined self-consistently from variational principle. For translation invariant system, the order parameter can be block diagonalized into $4\times 4$ momentum-independent matrices:
\begin{equation}
    P_{ki,k'j}^{0}=P_{ij}^{0}\, \delta_{k,k'}.
\end{equation}
As a result, the mean-field quasi-particle excitation is independent of $k$ and the 4 energy levels are obtained by diagonalizing the following mean-field Hamiltonian:
\begin{subequations}
\begin{align}
&\hat{h}=-\frac{\Delta_z}{2}s^{z}-\frac{\Delta_{v}}{2}\tau^{z}+\Sigma^{H}[\hat{P}]+\Sigma^{F}[\hat{P}],\label{def_HomogeneousH}\\
&\Sigma^{H}[\hat{P}]=\sum_{i=x,y,z}u_{i}\text{tr}(\tau^{i}\hat{P})\tau^{i},\ \ \Sigma^{F}[\hat{P}]=-u_{0}\hat{P}-\sum_{i=x,y,z}\tau^{i}\hat{P}\tau^{i}
\end{align}
\end{subequations}
where $u_{0}=\int\frac{d\vec{q}}{(2\pi)^2}U_{0}(\vec{q})e^{-{\vec{q}^2l_{B}^2}/{2}}$ is the exchange energy. Note that in this subsection we focus on the ground state and therefore omit the superscript $"0"$ for the simplicity of notation. The following two principles are useful guidelines to guess the correct ground state order parameter $P_{ij}$:

\begin{enumerate}
    \item  The exchange-energy of the dominant long-range Coulomb interaction ($u_0$) favors maximum isospin polarization, i.e.~Quantum Hall ferromagnetsim.

    \item If the short-range valley anisotropic interaction is a delta-function contact interaction $\delta(\vec{r}_i-\vec{r}_j)$, it does not scatter states with the same isospin due to Pauli exclusion principle.
\end{enumerate}
 
\textit{$\nu=\pm 1$}-- At filling factor $\nu=\pm1$ one of the four bands is empty (fill). Due to principle 1, electrons (holes) will occupy the band with identical isospin polarization. Then, the ``direction'' of the isopsin is solely selected by single-particle term and short-range Coulomb interactions does not play any role because of principle 2. Thus, the ground state of $\nu=\pm1$ is spin and valley polarized with the order parameter 
\begin{equation}
    P_{\nu=-1}=\ket{K\uparrow}\bra{K\uparrow}\;\;, \; \; P_{\nu=1}=\mathbf{1}-\ket{K'\downarrow}\bra{K'\downarrow}
\end{equation} 

\begin{figure*}
\includegraphics[width=0.9\columnwidth]{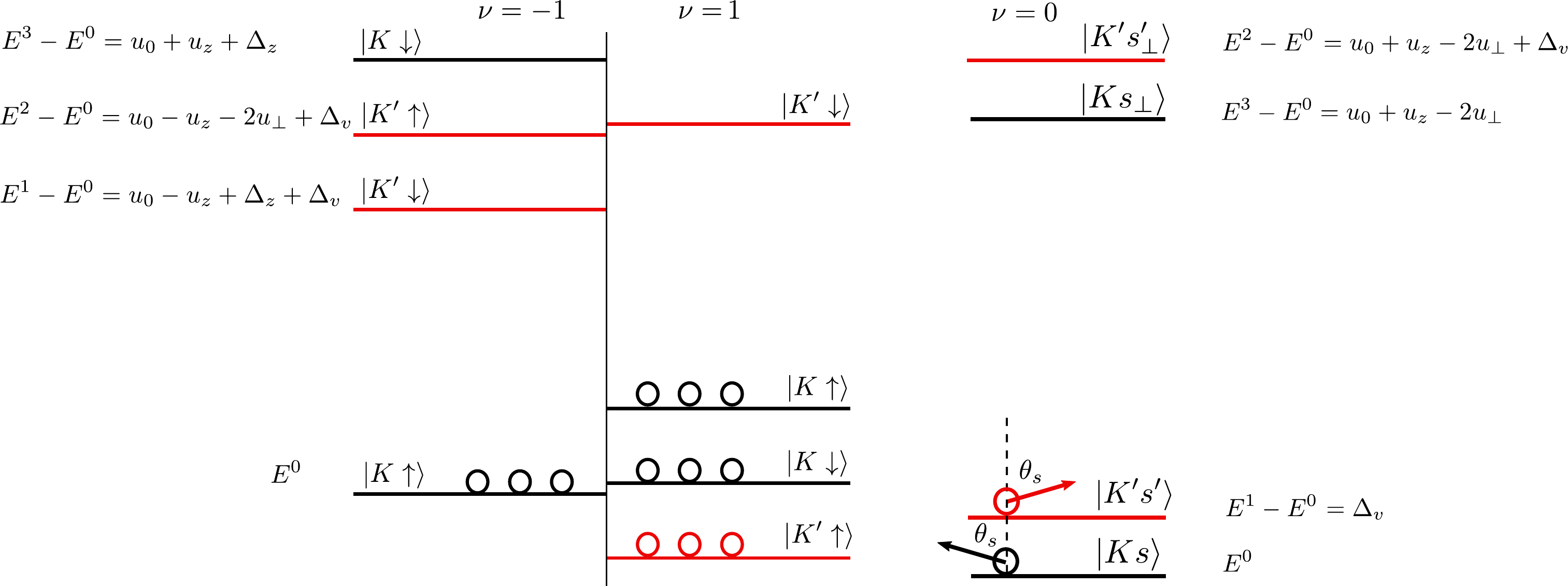}
\caption{A schematic of the energy levels and eigenstates of $\nu=\pm 1$ and $\nu=0$ QHMs. The expression of $E^{n}$ measured from the lowest level are provided for $\nu=-1,0$ QHMs, while the level spacing in $\nu=1$ QHM can be derived from $\nu=-1$ by a particle-hole transformation. Both of $\nu=\pm 1$ QHMs are assumed to be fully valley-polarized by a weak valley polarization energy $\Delta_{v}$. For $\nu=0$, $\Delta_{v}$ splits the otherwise degenerate (un)occupied bands. The arrows in the occupied bands represent the spin orientation with canting angle $\theta_{s}$.}\label{fig:band}
\end{figure*}

Energy levels of quasiparticle excitation are shown in Fig.~\ref{fig:band}. Bearing in mind the energy scale Eq.~\eqref{eq:heir} and experimental observation $-2u_\perp > \Delta_z$ in perpendicular magnetic field , we can understood the excited state ordering as follow. The three excited states are all separated from the ground-state by $u_0$ due to reversal of isospin. For a sample without sublattice polarization $\Delta_v=0$, excited states that flip valley polarization will require less energy because of Eq.~\eqref{eq:heir}. When $-2u_\perp > \Delta_z$, the first excited state flips both spin and valley while the second excited state only flips valley. This band ordering excitation can be experimentally adjust by a sublattice polarization potential $\Delta_v$ and in-plane magnetic field. 
Due to particle-hole symmetry, we obtain the band ordering of $\nu=1$ state by flipping the band ordering of $\nu=-1$ state and interchanging the isospin $K\leftrightarrow K'$ and $\uparrow\leftrightarrow \downarrow$ \cite{abanin2013fractional}. Before moving to the $\nu=0$ case, let us mention that when $\Delta_{v}=0$ the ground state at $\nu=\pm 1$ is subtle because the many-body Hamiltonian projected into the subspace of the valley-polarized states have a valley SU(2) symmetry, indicating that the system has to spontaneously choose a valley polarization. Finite temperature and disorder effect \cite{abanin2013fractional} will play an important role in this case.
Note however, when $\nu=\pm 1$ state forms a junction the valley degeneracy will be lifted \cite{wei2020band} and we can safely assume the ground state at $\nu=\pm 1$ is polarized in $K$ or $K'$ valley.
 
\textit{$\nu=0$}-- The charge neutral ($\nu=0$) state has to fill two out of the four $N=0$ Landau levels. Because the two occupied states have to be orthogonal to each other principle 1 does not select the ground state. If there were no single particle terms, 
the two occupied states will be polarized in opposite valley to minimize the $u_{z}$ self energy, and the spin in the two valley will polarize in opposite direction to minimize the $-u_{\perp}>0$ self-energy.

Since valley and sublattice are locked in $N=0$ Landau level, this means the ground state is an antiferromagnet. In the presence of finite Zeeman term, spins in opposite valley will cant towards the direction of total magnetic field and the ground state becomes a canted antiferromagnet (CAF), see Ref.~\cite{kharitonov2012phase} for more discussion. The order parameter of CAF is given by $P_{\nu=0}=1/2+\cos\theta_{s}s^{z}/2+\sin\theta_{s}\tau^{z}(\cos\phi s^{x}+\sin\phi s^{y})/2$ where $\theta_{s}$ is the canting angle satisfying $\cos\theta_{s}=\Delta_{z}/4|u_{\perp}|$. Due to the $U(1)$ symmetry of the spin rotation about $z$ axis, we take $\phi=0$. Physically, it means that the N\'{e}el vector of CAF state $\vec{l}=\text{tr}(\tau^{z}\vec{s}P)/2$ spontaneously polarizes in the $x$ direction. 

The energy levels and excitation spectrum of $\nu=0$ is shown in Fig.~1b. The four eigenvectors are $z^{n}$: 
\begin{align}
\begin{matrix}
z^{0}=\ket{Ks},&z^{1}=\ket{K's'},&z^{2}=\ket{K's_{\perp}'},&z^{3}=\ket{Ks_{\perp}}.
\end{matrix}\label{eq_eigenstates}
\end{align}
where $\ket{s}/\ket{s'}=(\cos\frac{\theta_{s}}{2},\pm\sin\frac{\theta_{s}}{2})^{T}$ and $\ket{s_{\perp}}/\ket{s_{\perp}'}=(\sin\frac{\theta_{s}}{2},\mp\cos\frac{\theta_{s}}{2})^{T}$.  Due to spin-canting, the excited state energy is independent of Zeeman energy.

\subsection{The RPA kernel and the normal modes of Quantum Hall Magnets}\label{sec:RPA}

The excitations for both $\nu=0$ and $\nu=\pm1$ we discussed so far are quasiparticle charge excitation. In this section, we discuss neutral collective excitation that typically occurs at much smaller energy scale.
In the main text, we have already sketched the derivation of the RPA equation. The expression of RPA kernel reads
\begin{align}
&\mathbb{K}_{kmn}^{k'm'n'}(q_{y})=(E_{k+q_{y}}^{m}-E_{k}^{n})\delta_{kk'}\delta_{mm'}\delta_{nn'}+(f_{n,k}-f_{m,k})\left(\tilde{V}_{mnn'm'}(k,k',q_{y})-\tilde{V}_{mm'n'n}(k'+q_{y},k',k-k')\right),\\
&\tilde{V}_{mnn'm'}(k,k',q_{y})=\sum_{\alpha}V_{\alpha}(k-k',q_{y})z_{k}^{m\dagger}\tau^{\alpha}z_{k+q_{y}}^{n}z_{k'+q_{y}}^{n'\dagger}\tau^{\alpha}z_{k'}^{m'}
\end{align}
In homogeneous QHMs, the quasiparticle spinors $z_{k}^{m}\equiv z^{m}$, energy $E_{k}^{m}\equiv E^{m}$ and the Fermi-Dirac distribution $f_{m,k}\equiv f_{m}$ are independent of the momentum $k$. Due to translation invariance, the RPA kernel $\mathbb{K}_{kmn}^{k'm'n'}(q_{y})$ is a function of $\Delta k=k'-k$ so it can be block diagonalized by Fourier transformation:
\begin{align}
    &\mathbb{K}_{mn}^{m'n'}(\vec{q})\nonumber\\
   =&\frac{1}{L_{y}}\sum_{\Delta k}\mathbb{K}_{k-\frac{\Delta k}{2}mn}^{k+\frac{\Delta k}{2} m'n'}(q_{y})e^{-iq_{x}\Delta k l_{B}^2}\nonumber\\
   =&(E^{m}-E^{n})\delta_{mm'}\delta_{nn'}+(f_{n}-f_{m})\left[\sum_{i=x,y,z}u_{i}(\vec{q})z^{m\dagger}\tau^{\alpha}z^{n}z^{n'\dagger}\tau^{\alpha}z^{m'}-\!\sum_{\alpha=0,x,y,z}u_{\alpha}(\vec{q})z^{m\dagger}\tau^{\alpha}z^{m'}z^{n'\dagger}\tau^{\alpha}z^{n}\right]\label{def_kmatrix},
\end{align}
where $u_{\alpha}(\vec{q})=\int\frac{d\vec{k}}{(2\pi)^2}U_{\alpha}(\vec{k})e^{-\vec{k}^2l_{B}^2/2}e^{-i\vec{q}\cdot\vec{k}l_{B}^2}$ only depends on the magnitude of the wave vector $\vec{q}$. In particular, $u_{i}(\vec{q})=u_{i}e^{-\vec{q}^2l_B^2/2}$ and in the $\kappa=0$ limit $u_{0}(\vec{q})=u_0e^{-\vec{q}^2l_B^2/4}I_0(\vec{q}^2l_B^2/4)$ where $u_{0}=\sqrt{\pi/2}e^2/\epsilon l_{B}$ and $I_{0}(x)$ is the modified Bessel function of the first kind. The particle-hole pairs $(mn)$'s fulfill $f_{m}\neq f_{n}$. Therefore, the isotropic($\alpha=0$) Hartree term does not appear in the first summation in the last line since $z^{m\dagger}z^{n}=0$. As a result, the RPA equation in the main text can be simplified as 
\begin{equation}
\omega(\vec{q})\phi_{mn}(\vec{q})=\sum_{m'n'}\mathbb{K}_{mn}^{m'n'}(\vec{q})\phi_{m'n'}(\vec{q})\label{eq_homoRPA},
\end{equation}
where by definition
\begin{equation}
\phi_{mn}(\vec{q})=\frac{1}{L_{y}}\sum_{k}\psi_{k-\frac{q_{y}}{2}mn}(q_{y},\omega(\vec{q}))e^{-iq_{x}kl_{B}^2}\label{eq_phi} 
\end{equation}
represents a normal mode of the density matrix fluctuation. One can explicitly verify that $\tilde{\phi}_{mn}=\phi_{nm}^{*}(-\vec{q})$ is another normal mode with frequency $-\omega(\vec{q})$ from Eq.~\eqref{eq_homoRPA} and $\mathbb{K}_{mn}^{m'n'}=-(\mathbb{K}_{nm}^{n'm'})^{*}$. Both modes must coexist in the system to preserve the hermicity of the density matrix $\hat{P}(t)$. To avoid the redundancy, we only focus on the positive-frequency modes in the following.

\textit{$\nu=0$ collective modes}-- The $\nu=0$ state has \textit{eight} particle-hole(p-h) transitions $\phi_{mn}$ which we labelled by $(mn)=(30),(21),(03)$, $(12),(20),(02),(31),(13)$. The index $m$ ($n$) represents the unoccupied (occupied) band. From Fig.~1b), we see that the first four pairs are intravalley excitations, while the last four are intervalley excitations. Using them as a basis to construct the matrix $\hat{\mathbb{K}}(\vec{q})$ in Eq.~\eqref{def_kmatrix}, we found it can be block diagonalized into two $4\times 4$ matrices,
\begin{equation}
\hat{\mathbb{K}}(\vec{q})=
\begin{pmatrix}
\hat{\mathbb{K}}^{\text{intra}}(\vec{q})&0\\
0&\hat{\mathbb{K}}^{\text{inter}}(\vec{q})
\end{pmatrix}.
\end{equation}
where the RPA kernel in intravalley and intervalley subspace is given by the following:
\begin{subequations}
\begin{align}
&\hat{\mathbb{K}}^{\text{inter}}(\vec{q})=
\begin{pmatrix}
\hat{\mathbb{K}}^{3}(\vec{q})&0\\
0&\hat{\mathbb{K}}^{4}(\vec{q})
\end{pmatrix},\\
    &\hat{\mathbb{K}}^{3,4}(\vec{q})=
    \begin{pmatrix}
    \Delta_{g}-u_{0}(\vec{q})+u_{z}(\vec{q})+2u_{\perp}(\vec{q})\sin^2\theta_{s}\pm\Delta_{v}&2u_{\perp}(\vec{q})\sin^2\theta_{s}\\
    -2u_{\perp}(\vec{q})\sin^2\theta_{s}&-\left(\Delta_{g}-u_{0}(\vec{q})+u_{z}(\vec{q})+2u_{\perp}(\vec{q})\sin^2\theta_{s}\pm\Delta_{v}\right)\\
    \end{pmatrix}.
\end{align}
\end{subequations}   
\begin{equation}
\hat{\mathbb{K}}^{\text{intra}}(\vec{q})=
\begin{pmatrix}
\Delta_{g}-u_{0}(\vec{q})-u_{z}(\vec{q})&2u_{\perp}(\vec{q})\cos^2\theta_{s}&-2u_{\perp}(\vec{q})\sin^2\theta_{s}&0\\
    2u_{\perp}(\vec{q})\cos^2\theta_{s}&\Delta_{g}-u_{0}(\vec{q})-u_{z}(\vec{q})&0&-2u_{\perp}(\vec{q})\sin^2\theta_{s}\\
    2u_{\perp}(\vec{q})\sin^2\theta_{s}&0&-(\Delta_{g}-u_{0}(\vec{q})-u_{z}(\vec{q}))&-2u_{\perp}(\vec{q})\cos^2\theta_{s}\\
    0&2u_{\perp}(\vec{q})\sin^2\theta_{s}&-2u_{\perp}(\vec{q})\cos^2\theta_{s}&-(\Delta_{g}-u_{0}(\vec{q})-u_{z}(\vec{q}))
\end{pmatrix},\label{eq_kintra}
\end{equation}
where $\Delta_{g}=u_{0}+u_{z}-2u_{\perp}$ is the bulk gap of CAF state without any sublattice polarization potential, $\Delta_{v}=0$.

The eigenvalues of $\hat{\mathbb{K}}^{\text{inter}}$ and $\hat{\mathbb{K}}^{\text{intra}}$ are the intervalley and intravalley collective mode dispersion. Their dispersion is documented in Table.~\ref{tab:dispersion}. We found the intervalley modes have large excitation gaps $\omega_{3,4}(\vec{q}=0) \sim 11 \Delta_z \mp \Delta_v$  so they do not contribute to nonlocal spin-transport experiments which typically occurs at energy scale $\lesssim 5\Delta_z$.  
In contrast, intravalley modes, namely the (gapless) N\'{e}el mode $\phi^{1}$ and the Larmor mode $\phi^{2}$ play a significant role in nonlocal spin transport experiments. In the basis $\{(30),(21),(03),(12)\}$, their wave functions are given by the following in
\begin{subequations}
\label{eq_wavefunction}
\begin{align}
     &\phi^{\alpha}(\vec{q})=(u_{\vec{q}}^{\alpha }\, , \,(-1)^{\alpha-1}u_{\vec{q}}^{\alpha} \,, \,v_{\vec{q}}^{\alpha}\, ,\, (-1)^{\alpha-1}v_{\vec{q}}^{\alpha})^{T}\\
     &u_{\vec{q}}^{\alpha }=\frac{1}{2}\sqrt{1+\frac{\xi_{\alpha}(\vec{q})}{\omega_{\alpha}(\vec{q})}},\quad v_{\vec{q}}^{\alpha}=\frac{1}{2}\sqrt{-1+\frac{\xi_{\alpha}(\vec{q})}{\omega_{\alpha}(\vec{q})}},\\
     &\xi_{\alpha}(\vec{q})=\Delta_{g}-u_{0}(\vec{q})+u_{z}(\vec{q})+(-1)^{\alpha}u_{\perp}(\vec{q})\cos^2\theta_{s},\  \alpha=1,2.
\end{align}
\end{subequations}%
It is easy to verify that $(u_{\vec{q}}^{\alpha})^2-(v_{\vec{q}}^{\alpha})^2=1/2$ and hence the wave functions satisfy the normalization condition
\begin{equation}
\sum_{mn}(f_{m}-f_{n})\bar{\phi}_{mn}^{\alpha}(\vec{q})\phi_{mn}^{\beta}(\vec{q})=\delta^{\alpha\beta}.~\label{eq_normalization}
\end{equation}
where $\bar{\phi}$ is the complex conjugate of $\phi$. 
We note the Kernel matrix $\hat{\mathbb{K}}^{\text{intra}}(\vec{q})$ has an addition $\mathbb{Z}_{2}$ symmetry $\mathbf{1}\otimes\rho^{x}$ such that $[\mathbf{1}\otimes\rho^{x},\hat{\mathbb{K}}^{\text{intra}}(\vec{q})]=0$, where $\rho^{x}$ is the 1st Pauli matrix. One can easily check $\mathbf{1}\otimes\rho^{x}\phi^{1,2}(\vec{q})=\pm\phi^{1,2}(\vec{q})$. Therefore, the gapless  N\'{e}el mode and the Larmor mode with Zeeman gap are repsectively the symmetric and anti-symmetric combinations of the spin-flipping excitation in two valleys (or sublattice).

In order to understand dynamics of the observables,
 we first use the excited state wavefunctions of the RPA equation to construct the fluctuating density matrix: 

\begin{align}\label{eq_rho}
    \rho_{ij}(\vec{r},t)&=\bra{\Psi_{\text{QHM}}(t)}\hat{\psi}_{j}^{\dagger}(\vec{r})\hat{\psi}_{i}(\vec{r})\ket{\Psi_{\text{QHM}}(t)}\nonumber\\
    &=\frac{1}{L_{y}}\sum_{k,q_{y}}\frac{1}{\sqrt{\pi}l_{B}}e^{-\frac{(x-kl_{B}^2)^2}{l_{B}^2}-\frac{q_{y}^2l_{B}^2}{4}}e^{iq_{y}y}P_{k+\frac{q_{y}}{2}i,k-\frac{q_{y}}{2}j}(t)\nonumber\\
    &=P_{ij}^{0}+\sum_{l=1,2}\int \left[
    \delta P_{ij}^{l}(\vec{q})e^{i(\vec{q}\cdot\vec{r}-\omega_{l}(\vec{q})t)}
    + \left(\delta P_{ji}^{l}(\vec{q})\right)^{*}e^{-i(\vec{q}\cdot\vec{r}-\omega_{l}(\vec{q})t)}
    \right] \frac{d\vec{q}}{(2\pi)^{2}}
\end{align}
where we have expanded the density matrix to linear order in deviation
\begin{equation}
    P_{k+\frac{q_{y}}{2}i,k-\frac{q_{y}}{2}j}(t)=\bra{\Psi_{\text{QHM}}(t)}c_{k-\frac{q_{y}}{2}j}^{\dagger}c_{k+\frac{q_{y}}{2}i}\ket{\Psi_{\text{QHM}}(t)}=P_{ij}^{0}\delta_{q_{y},0}+\delta P_{k+\frac{q_{y}}{2}i,k-\frac{q_{y}}{2}j}(t)+O(\delta P^2)
\end{equation}
here $P_{ij}^{0}$ is the ground state order parameter discussed in Sec.~\ref{sec:meanfield}. The integral in the last line of Eq.~\eqref{eq_rho} is obtained by expanding $\delta \hat{P}(t)$ in terms of normal modes. The first and second integrands correspond to positve-and negative-frequency modes, respectively. Comparing with the definition of the normal mode wave function, Eq.~\eqref{eq_phi}, we arrive at the following relation,
\begin{equation}
    \delta P_{ij}^{l}(\vec{q})=a^{l}(\vec{q})\sum_{mn}\phi_{mn}^{l}(\vec{q})z_{i}^{m}z_{j}^{n\dagger}  \label{eq_eigenP}
\end{equation}
where the small parameter $a^{l}$ denotes the amplitude of the $l$th normal mode. Substituting the $\nu=0$ quasiparticle spinors Eq.~\eqref{eq_eigenstates} into the above equation and using the long-wavelength limit of the dispersion,

\begin{align}
&\omega_{1}(\vec{q})\approx v_{\mathrm{AF}}|\vec{q}|,\ \ \ v_{\mathrm{AF}}l_{B}^{-1}=\sqrt{\left[u_{0}+2u_{z}+4|u_{\perp}|\right]|u_{\perp}|\sin^2\theta_{s}},
\end{align}
we arrive at the following:
\begin{subequations}
\label{eq_P12}
\begin{equation}
    \delta\hat{P}^{1}(\vec{q})/a^{1}(\vec{q})=i\tau^{z}s^{y}+\frac{v_{\mathrm{AF}}|\vec{q}|}{4|u_{\perp}|\sin^2\theta_{s}}(\sin\theta_{s} s^z-\cos\theta_{s}\tau^{z}s^{x}+i\tau^{z}s^{y})+O(\vec{q}^2l_{B}^2)\label{eq_P1},
\end{equation}
\begin{equation}
    \delta \hat{P}^{2}(\vec{q})/a^{2}(\vec{q})=-\cos\theta_{s}(s^{x}-is^{y})+\sin\theta_{s}\tau^{z}s^{z}+O(\vec{q}^2l_{B}^2)\label{eq_P2}.
\end{equation}
\end{subequations}


At $\vec{q}=0$, $\phi^1$ or Eq.~\eqref{eq_P1} describes a global rotation of N\'{e}el vector which costs zero energy, see Table.~\ref{tab:dispersion}. It disperses linearly at finite $q$ and in addition to the fluctuation of azimuthal angle of the N\'{e}el vector, it also generates fluctuation of spin-density along the broken symmetry direction, i.e.~$z$. In Eq.~\eqref{eq_P2}, the first term term describes precession of total spin about the $z$ axis ($s^x-is^y=s^-$ is a spin-lowering operator) so this corresponds to the Larmor mode that has an energy gap of Zeeman energy, see Table.~1. Because the N\'{e}el vector has to be perpendicular to total spin-polarization locally ($\vec{s}\cdot \vec{l}=0$), the Larmor mode will also tilt the  N\'{e}el vector towards to $z$ direction and this is describes by the second term in Eq.~\eqref{eq_P2}.


\textit{$\nu=1$ collective modes}-- Let us label the particle-hole excitation of $\nu=1$ mode by the compound index $(mn)=(10),(20),(30),(01),(02),(03)$. 
They constitute a basis under which the $6\times 6$ matrix $\hat{\mathbb{K}}(\vec{q})$ is diagonal. The first three excitations have positive frequencies and are listed in Table.~\ref{tab:dispersion}. The corresponding normal modes are
\begin{equation}
\phi^{1}=(1,0,0,...)^{T},\ \phi^{2}=(0,1,0,...)^{T},\ \phi^{3}=(0,0,1,0,...)^{T}.
\end{equation}
$\nu=-1$ QHM has the same collective mode dispersion due to the particle-hole symmetry.
Among all these collective modes, we mainly focus on the magnons in the main text. From Table.~\ref{tab:dispersion}, we derive the long-wave length apprximation of the magnon dispersions,
\begin{align}
&\omega_{s}(\vec{q})=\Delta_{z}+2\rho_{s}\vec{q}^2,\ \ \ \rho_{s}l_{B}^{-2}=\frac{u_{0}}{8}+\frac{u_{z}}{4}.
\end{align}
When $\frac{\Delta_{z}}{4|u_{\perp|}}\ll 1$, the $\nu=\pm 1$ magnons have lower energy than $\nu=0$ magnon,  $\omega=\omega_{s}(\vec{q})<\omega_{1}(\vec{q})$, except for $\omega<\Delta_{z}(1+2\rho_{s}\Delta_{z}/v_{\mathrm{AF}}^{2})$, which is however merely a narrow range because
\begin{equation}
\frac{2\rho_{s}\Delta_{z}}{v_{\mathrm{AF}}^{2}}=\frac{(\frac{u_{0}}{4}+\frac{u_{z}}{2})\Delta_{z}}{(u_{0}+2u_{z}+4|u_{\perp}|)|u_{\perp}|}<\frac{\Delta_{z}}{4|u_{\perp}|}.
\end{equation}
Note that we used $\sin\theta_{s}\approx 1$ to simplify the analysis.
\begin{table}
    \centering
    \caption{A list of collective mode dispersion of $\nu=0$ CAF phase and $\nu=1$ valley-and-spin-polarized QHM.}
    \begin{ruledtabular}
    \begin{tabular}{l l l}
    $\nu$&\textrm{collective mode}&\textrm{dispersion}\\
    \hline
    $0$&gapless mode $\phi^{1}$/Larmor mode $\phi^{2}$&$\omega_{1,2}=\sqrt{\left[\Delta_{g}-u_{0}(\vec{q})-u_{z}(\vec{q})\pm 2u_{\perp}(\vec{q})\cos^2\theta_{s}\right]^2-4u_{\perp}^2(\vec{q})\sin^4\theta_{s}}$\\
    &intervalley mode $\phi^{3,4}$&$\omega_{3,4}=\sqrt{(\Delta_{g}-u_{0}(\vec{q})+u_{z}(\vec{q}))(\Delta_{g}-u_{0}(\vec{q})+u_{z}(\vec{q})+4u_{\perp}(\vec{q})\sin^2\theta_{s})}\mp\Delta_{v}$\\
    $1$&spin wavey(magnon) $\phi^{1}$&$\omega_{s}=u_{0}-u_{0}(\vec{q})+u_{z}-u_{z}(\vec{q})+\Delta_{z}$\\
    &valley wave $\phi^{2}$&$\omega_{v}=u_{0}-u_{0}(\vec{q})-(u_{z}-u_{z}(\vec{q}))-2(u_{\perp}-u_{\perp}(\vec{q}))+\Delta_{v}$\\
    &spin-valley wave $\phi^{3}$&$\omega_{vs}=u_{0}-u_{0}(\vec{q})-(u_{z}-u_{z}(\vec{q}))+\Delta_{z}+\Delta_{v}$
    \end{tabular}\\
    \end{ruledtabular}
    \label{tab:dispersion}
\end{table}

\section{Numerical Method to Calculate the S-matrix of collective modes in Quantum Hall magnet (QHM) junctions}

In this section, we describe a numerical method to calculate the transmission probability of collective modes from the RPA equation. For simplicity, we focus on a one-dimensional scattering problem defined in the $x$ direction and apply periodic boundary condition in the $y$-direction. In the Landau gauge, the momentum $k$ describing the plane-wave along the $y$-direction also means the wavefunction is localized at the guiding center coordinate $X=kl_B^2$. The guiding centers are equally spaced $X_{i}-X_{i-1}=2\pi/L_{y}$ in a system with fixed width $L_y$. Let the scattering geometry (i.e.~QHM junction) be described by a set of guiding centers $\{X_{i}|i=1,..,N\}$ and we study the transmission probability of an incoming collective mode in $X<X_1$ to an outgoing collective mode in $X>X_N$. 
Recall in the maintext, we use a compact notation $\psi_{kmn}(q_{y},\omega)$ to describe particle-hole transition between band $m$ and $n$ with transverse momentum $q_y$ and frequency $\omega$. The collective modes of the homogeneous QHM are the asymptotic states of the scattering problem.

From here and what follows,  we use $k=Xl_B^2$ interchangeably and the superscripts $\alpha,\beta$ and $(\alpha',\beta')$ to label collective modes in $X<X_1$ and $X>X_N$ region. They are given by the following:

\begin{equation} 
\psi_{Xmn}(q_{y},\omega)=
\begin{cases}
\frac{1}{\sqrt{v_{\alpha}}}\phi_{mn}^{\alpha}(q_{x}^{\alpha},q_{y})e^{iq_{x}^{\alpha}X}+\sum\limits_{\beta}r_{\beta\alpha}\frac{1}{\sqrt{v_{\beta}}}\phi_{mn}^{\beta}(-q_{x}^{\beta},q_{y})e^{-iq_{x}^{\beta}X},\ &X\leq X_{1}\\
\sum\limits_{\beta'}t_{\beta'\alpha}\frac{1}{\sqrt{v_{\beta'}}}\phi_{mn}^{\beta'}(q_{x}^{\beta'},q_{y})e^{iq_{x}^{\beta'}X},&X\geq X_{N}
\end{cases}\label{eq_bc}
\end{equation}
where the normal component of the wave vector $q_{x}^{\alpha}$ is a positive solution to the following equation,
\begin{equation} 
\omega_{\alpha}(q_{x}^{\alpha},q_{y})=\omega,\label{def_qx}
\end{equation}
and $v_{\alpha}=(dq_{x}^{\alpha}/d\omega)^{-1}$ is the velocity of the collective mode. 

The unknown parameters $r$ and $t$ in Eq.~\eqref{eq_bc} can be eliminated using the normalization condition Eq.~\eqref{eq_normalization} and the wavefunction at the start of the junction $X=X_1$ and end of the junction $X=X_N$:
\begin{subequations}
\label{eq_rt}
\begin{align}
&r_{\beta\alpha}=\left[-\delta_{\alpha\beta}e^{iq_{x}^{\alpha}X_{1}}+\sqrt{v_{\beta}}\sum_{mn}(f_n-f_m)\bar{\phi}_{mn}^{\beta}(-q_{x}^{\beta},q_{y})\psi_{X_{1}mn}\right]e^{iq_{x}^{\beta}X_{1}},\\
&t_{\beta'\alpha}=\sqrt{v_{\beta'}}\sum_{mn}(f_n-f_m)\bar{\phi}_{mn}^{\beta'}(q_{x}^{\beta'},q_{y})\psi_{X_{N}mn}e^{-iq_{x}^{\beta'}X_{N}}.
\end{align}
\end{subequations}

Next, we substitute Eq.~\eqref{eq_rt} and Eq.~\eqref{eq_bc} into the RPA equation in the main text, we arrive at the main equation to be solved numerically:
\begin{equation}
\sum_{i'=1}^{N}\sum_{m'n'}\left[(\mathbb{K}^{\text{eff}})_{X_{i}mn}^{X_{i'}m'n'}(q_{y},\omega)-\omega\delta_{ii'}\delta_{mm'}\delta_{nn'}\right]\psi_{X_{i'}m'n'}(q_{y},\omega)=V_{X_{i}mn}^{\alpha}(q_{y},\omega),\ \ \ 1\leq i\leq N\label{eq_modifiedRPA}
\end{equation}
Eq.~\eqref{eq_modifiedRPA} is the RPA equation of maintext (Eq.~5) expressed in a finite domain where states to the left $X\leq X_1$ and to the right $X \geq X_N$ are fixed. The effective RPA Kernel $\hat{\mathbb{K}}^{\text{eff}}$ accounts for the (super) self-energy from states in $X<X_1$ and $X>X_N$:


%
\begin{subequations}
\begin{align}
&\hat{\mathbb{K}}^{\text{eff}}(q_{y},\omega)=\hat{\mathbb{K}}(q_{y})+\hat{\Sigma}^{L}(q_{y},\omega)+\hat{\Sigma}^{R}(q_{y},\omega),\\
&(\Sigma^{^{L}})_{Xmn}^{X'm'n'}(q_{y},\omega)=\delta_{X',X_{1}}(f_{n}-f_{m})\sum_{\beta}\sum_{j<1}\sum_{\bar{m}\bar{n}}\mathbb{K}_{Xmn}^{X_{j}\bar{m}\bar{n}}(q_{y})e^{-iq_{x}^{\beta}(X_{j}-X_{1})}\phi_{\bar{m}\bar{n}}^{\beta}(-q_{x}^{\beta},q_{y})\bar{\phi}_{mn}^{\beta}(-q_{x}^{\beta},q_{y}),\\
&(\Sigma^{R})_{Xmn}^{X'm'n'}(q_{y},\omega)=\delta_{X',X_{N}}(f_{n}-f_{m})\sum_{\beta'}\sum_{j>N}\sum_{\bar{m}\bar{n}}\mathbb{K}_{Xmn}^{X_{j}\bar{m}\bar{n}}(q_{y})e^{iq_{x}^{\beta'}(X_{j}-X_{N})}\phi_{\bar{m}\bar{n}}^{\beta'}(q_{x}^{\beta'},q_{y})\bar{\phi}_{mn}^{\beta'}(q_{x}^{\beta'},q_{y});
\end{align}
\end{subequations}
In addition to the renormalization of RPA kernel, the incoming wave also introduces a source term in the RHS of Eq.~\eqref{eq_modifiedRPA}:

\begin{equation}
V_{Xmn}^{\alpha}(q_{y},\omega)=\delta_{X,X_{1}}\frac{1}{\sqrt{v_{\alpha}}}\phi_{mn}^{\alpha}(-q_{x}^{\alpha},q_{y})\sum_{i'<1}e^{-iq_{x}^{\alpha}(X_{i'}-X_{1})}-e^{iq_{x}^{\alpha}X_{i'}}.
\end{equation}
Note that for given $(q_{y},\omega)$, $(\mathbb{K}^{\textrm{eff}})_{Xmn}^{X'm'n'}$ and $V_{Xmn}^{\alpha}$ are fully determined without any unknown parameters, so the linear equation Eq.~\eqref{eq_modifiedRPA} can be solved straightforwardly. From the output wave function $\psi_{Xmn}$, we can read out $r_{\alpha\beta}$ and $t_{\alpha\beta'}$ from Eqs.~\eqref{eq_rt}.

Similarly, if a collective mode $\phi^{\alpha'}(\vec{q})$ is injected from the right, the asymptotic wave function reads that
\begin{equation}
\psi_{Xmn}(q_{y},\omega)=
\begin{cases}
\sum\limits_{\beta}t_{\beta\alpha'}'\frac{1}{\sqrt{v_{\beta}}}\phi_{mn}^{\beta}(q_{x}^{\beta},q_{y})e^{-iq_{x}^{\beta}X},\ &X\leq X_{1}\\
\frac{1}{\sqrt{v_{\alpha'}}}\phi_{mn}^{\alpha'}(q_{x}^{\alpha'},q_{y})e^{-iq_{x}^{\alpha'}X}+\sum\limits_{\beta'}r_{\beta'\alpha'}'\frac{1}{\sqrt{v_{\beta'}}}\phi_{mn}^{\beta'}(q_{x}^{\beta'},q_{y})e^{iq_{x}^{\beta'}X},&X\geq X_{N}\\
\end{cases}\label{eq_bc2}
\end{equation}
Following a similar procedure, we can evaluate $r_{\alpha\beta}'$ and $t_{\alpha\beta'}'$. The S-matrix of the QHM junction is constructed as follows:
\begin{equation}
S(q_{y},\omega)=
\Biggl(\mkern-5mu
\begin{tikzpicture}[baseline=-.65ex]
\matrix[
  matrix of math nodes,
  column sep=1ex,
] (m)
{
r_{\beta\alpha} & t'_{\beta\alpha'} \\
t_{\beta'\alpha} & r'_{\beta'\alpha'} \\
};
\draw[dotted]
  ([xshift=0.5ex]m-1-1.north east) -- ([xshift=0.5ex]m-2-1.south east);
\draw[dotted]
  (m-1-1.south west) -- (m-1-2.south east);
\end{tikzpicture}\mkern-5mu
\Biggr).
\end{equation}
In the main text, we studied the magnon scattering problem in $\nu=1|\nu_{m}|1$ QHM junctions. Because the microscopic Hamiltonian, Eq.~\eqref{eq_H}, conserves the total valley quantum number and the mean-field quasiparticle states do not mix different valleys (see Fig.1 in the maintext and Ref.~\cite{wei2020band}), the magnon, as an intravalley mode, is decoupled from intervalley excitations. Consequently, the S-matrix of the magnon at a given parallel momentum $q_{y}$ and energy $\omega$ is reduced to a $2\times 2$ matrix, 
$\begin{pmatrix}
r(q_{y},\omega)&t'(q_{y},\omega)\\
t(q_{y},\omega)&r'(q_{y},\omega)
\end{pmatrix}.$ 


\end{document}